\documentclass[%
 aip,
 amsmath,amssymb,
 reprint,%
]{revtex4-1}

\usepackage{graphicx}
\usepackage{dcolumn}
\usepackage{bm}

\usepackage[utf8]{inputenc}
\usepackage[T1]{fontenc}
\usepackage{mathptmx}
\usepackage{etoolbox}
\usepackage{color}

\makeatletter
\def\@email#1#2{%
 \endgroup
 \patchcmd{\titleblock@produce}
  {\frontmatter@RRAPformat}
  {\frontmatter@RRAPformat{\produce@RRAP{*#1\href{mailto:#2}{#2}}}\frontmatter@RRAPformat}
  {}{}
}%
\makeatother
\begin{document}

\preprint{AIP/123-QED}

\title[]{Drying-Induced  Size Segregation and Stratification in Colloidal Droplets across Binary, Ternary, and  Polydisperse Systems}
\author{Zhaoxuan Liu}
\author{Marjolein Dijkstra}%
 \email{m.dijkstra@uu.nl.}
\affiliation{ Soft Condensed Matter \& Biophysics group, Debye Institute for Nanomaterials Science, Utrecht University, Princetonplein 1, 3584 CC Utrecht, the Netherlands.}%

\date{\today}

\begin{abstract}
We study evaporation-induced size segregation and stratification in  binary, ternary, and polydisperse mixtures of colloidal particles within a drying droplet using Brownian dynamics (BD) simulations. We show that drying-induced stratification can occur in all three types of mixtures, with larger particles preferentially accumulating in the droplet interior and smaller particles near the droplet surface. In binary mixtures, size segregation becomes more pronounced as the Péclet number and the size difference between particles increase, and as the relative volume fraction of large particles and the initial particle concentration decrease.  
In ternary mixtures, three distinct layers form when the Péclet number is sufficiently high  and the intermediate particle size is suitably selected. Under the conditions studied, increasing the fraction of intermediate-sized particles can also enhance separation between the small and large particles. These results are in qualitative agreement with previous studies of drying planar films. Extending the analysis to polydisperse systems, we find that the average particle size decreases with increasing radial distance under stratifying conditions. 
Furthermore, higher Péclet numbers and particle-size distributions with more particles at both size extremes lead to steeper radial gradients in particle size. Additionally, we use multi-particle collision dynamics (MPCD) simulations to assess the role of hydrodynamic interactions. BD simulations predict stronger stratification than MPCD simulations, although stratification persists in the MPCD simulations. Our results thus demonstrate that droplet-confined drying offers a simple and versatile strategy for engineering radial composition and size gradients in colloidal assemblies, thereby offering new opportunities for the controlled fabrication of hierarchically structured supraparticles. 
\end{abstract}

\maketitle

\section{\label{intro}Introduction}

As a colloidal droplet dries, its receding interface concentrates the suspended particles in a progressively shrinking volume. In mixtures of different-sized particles, this process can produce nonuniform radial size distributions, ranging from smooth composition gradients to sharply defined layers. In this work, we examine how drying conditions and particle composition affect the resulting stratification patterns. 

We first consider stratification during the drying of planar films, a process  relevant to the formation of paint and cosmetic coatings\cite{keddie2010fundamentals,taylor2011wetting}. As the solvent evaporates, particles can redistribute across the film thickness~\cite{routh2004distribution}, shaping the structure and properties of the dried coating. In drying colloidal films containing both large and small particles, spontaneous stratification can arise from the interplay between particle diffusion and the receding interface. Large-on-top stratification occurs when the diffusive mobilities of the particles differ sufficiently. When the small particles diffuse much faster than the drying rate, while the large particles diffuse much more slowly~\cite{trueman2012auto, trueman2012autostratification}, the small particles can diffuse away from the descending interface. In contrast, the large particles lag behind and accumulate near the top surface.  Interestingly, small-on-top stratification can emerge when all particles diffuse slowly relative to  the drying rate~\cite{fortini2016dynamic,fortini2017stratification,sear2017diffusiophoresis,rees2021stratification}. The resulting behavior depends on the evaporation rate, particle size ratio, diffusion coefficients of the different species, film thickness, initial particle concentrations, and interparticle interactions~\cite{zhou2017cross, howard2017stratification,makepeace2017stratification,schulz2018critical,sear2018stratification,tang2018stratification,tang2019stratification, tinkler2022effect,hooiveld2025self}. Within the parameter ranges studied, free-draining models which neglect hydrodynamic interactions show that  faster drying and larger particle size ratios can promote small-on-top stratification.  Moreover, small-on-top stratification is not limited to binary mixtures; it has also been demonstrated to occur in ternary and even fully polydisperse colloidal systems~\cite{fortini2017stratification,tinkler2021evaporation}, as well as mixtures of spheres and anisotropic particles~\cite{liu2025effect}.

In a drying spherical droplet, the receding interface confines particles from all directions, causing stratification to develop radially rather than across the thickness of a planar film. When such droplets evaporate sufficiently slowly, ordered crystalline structures can emerge within these resulting supraparticles in both single-component and binary systems~\cite{de2015entropy, wang2021binary}. In contrast, under rapid drying conditions particles tend to self-stratify, forming a core-shell structure with large particles in the interior and small particles in the  outer shell~\cite{liu2019segregation,wolf2025numerical,jee2026stratification}.  This shell may subsequently buckle into a non-spherical supraparticle, as described by the theory of thin elastic shells~\cite{zhou2022tuning}. 
Particle shape represents an additional key control parameter in evaporation-driven self-assembly. Numerous studies have shown that nanoparticle assembly within drying emulsion droplets provides a versatile route to hierarchical superstructures. In these confined geometries, the interplay between particle shape, packing constraints, and spherical confinement governs both positional and orientational order. For example, sharp cubic nanoparticles assemble into simple-cubic lattices, whereas rounded cubes form icosahedral clusters in slowly drying droplets~\cite{wang2018interplay}. Similarly, the intrinsic phase behavior of nanoplatelets enables the simultaneous three-dimensional positional and orientational order under confinement~\cite{wang2022structural}.
In rapidly drying droplets containing mixtures of spheres and ellipsoids, shape anisotropy leads to spatial segregation. Experiments reveal the formation of an ellipsoid-rich surface layer, while simulations further indicate the presence of a  sphere-rich region immediately beneath this outer layer~\cite{yetkin2023structure}.

In this work, we investigate evaporation-induced self-stratification and size segregation in binary mixtures of colloidal spheres within a drying droplet using computer simulations. We first investigate how the drying rate, particle size ratio, initial particle concentration, and relative volume fraction of large particles shape the resulting spatial particle distributions in binary systems. We then extend the analysis to ternary and polydisperse mixtures. For ternary mixtures, we examine how the size and volume fraction of the intermediate species affect stratification. For polydisperse mixtures, we study how the initial particle-size distribution governs the radial variation in particle size within the drying droplet. Finally, we assess the effect of hydrodynamic interactions on  stratification.

By systematically investigating how drying conditions and particle composition influence stratification, we gain insight into the formation and internal structure of supraparticles. Such understanding is essential  for the rational design of functional supraparticles for  applications in catalysis\cite{li2017hierarchically,liu2019tuning,hou2019formation}, gas adsorption and sensing\cite{fujiwara2023fast,fang2009single}, drug encapsulation and delivery\cite{rial2014bioinspired, zhao2015microfluidic}, photonics\cite{manoharan2001photonic,park2014full,vogel2015color}, and energy conversion and storage\cite{koo2008nano,li2014mesoporous}.

This paper is organized as follows. In Section II, we describe the model and simulation methodology. Section III presents and discusses the results for binary, ternary and polydisperse systems, as well as the effects of hydrodynamic interactions. Finally, Section IV summarizes our conclusions.

\section{Model and simulation method}
\label{model}
\subsection{Brownian dynamics simulations}
\label{BD}
\begin{figure}[htbp]
\includegraphics{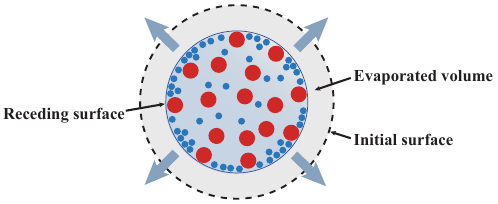}
\caption{\label{fig:illu1} Schematic illustration of the drying process in a spherical droplet, highlighting the evolution of particle distributions within a droplet upon solvent evaporation.}
\end{figure}

We consider a system  of $N$ spherical particles, labeled  $i=1,\dots,N$, each with diameter $\sigma_i$ at position $\mathbf{r}_i$, confined within a spherical droplet of radius $R(t)$. 
The droplet confinement is modeled as a soft spherical boundary whose radius decreases with time $t$, thereby mimicking the effect of evaporation-induced compression, as illustrated in Fig.~\ref{fig:illu1}. 
The time evolution of the droplet radius follows \cite{kundu2025exploring, langmuir1918evaporation}
\begin{eqnarray}
R(t) = \sqrt{R^2(0) - \alpha t},
\label{eq1}
\end{eqnarray}
where $R(0)$ denotes the initial droplet radius and $\alpha$ controls the rate at which the boundary shrinks. The instantaneous speed $v$ of the receding droplet interface is therefore
\begin{equation}
v(t)=-\frac{d R(t)}{dt} = \frac{\alpha}{2R(t)}, 
\end{equation}
where the negative derivative indicates that the radius decreases over time. The evaporation process can be characterized by the P\'eclet number ${ Pe}_i=v(0)R(0)/D_i=\alpha/(2D_i)$, which quantifies the relative importance of advective transport to diffusive motion of particle $i$. Here, $D_i$ denotes the diffusion coefficient of particle $i$.

The position $\mathbf{r}_i$ of particle $i$, measured relative to the center of the spherical confinement, evolves according to Brownian dynamics~\cite{ermak1978brownian}
\begin{eqnarray}
\mathbf{\dot r}_i(t)=
\frac{D_i}{k_BT}\mathbf{F}_i(t) +\boldsymbol{\eta}_i(t) ,
\label{eq2}
\end{eqnarray}
where $k_B$ is the Boltzmann constant, $T$ the temperature, and $\mathbf{F}_i$  the total force acting on particle $i$. The stochastic noise $\boldsymbol{\eta}_i$  acting on particle $i$ is  characterized by a  zero mean $\langle \eta_{i,\alpha}(t)\rangle=0$ and variance $\langle \eta_{i,\alpha}(t) \eta_{j,\beta}(t')\rangle=2 D_i\delta_{\alpha\beta}\delta_{ij}\delta(t-t')$, where $\alpha$ and $\beta$ denote Cartesian components. The total force $\mathbf{F}_i$ acting on particle $i$ includes contributions from  particle--particle  and particle--wall interactions
\begin{equation}
    \mathbf{F}_i=-\sum_{j\ne i}\nabla U_{pp}(r_{ij})-\nabla U_{pw}(r_i,t),
    \label{eqforce}
\end{equation}
where $r_i=|\mathbf{r}_i|$, $r_{ij} = |\mathbf{r}_{ij}| = |\mathbf{r}_i - \mathbf{r}_j|$ represents the center-to-center distance between particles $i$ and $j$, and $U_{pp}(\mathbf{r}_{ij})$ denotes the purely repulsive interaction energy, described by the Weeks-Chandler-Andersen (WCA) potential \cite{weeks1971role} 
\begin{eqnarray}
U_{pp}(r_{ij}) 
=\left\{
\begin{array}{ll}
4\epsilon \left[\left(\dfrac{\sigma_{ij}}{r_{ij}}\right)^{12}
\hspace{-3mm}- \left(\dfrac{\sigma_{ij}}{r_{ij}}\right)^6 \right] + \epsilon,
& r_{ij} \leq 2^{1/6}\sigma_{ij}, \\[6pt]
0, & r_{ij} > 2^{1/6}\sigma_{ij},
\end{array}
\right.
\label{eqWCA}
\end{eqnarray}
where $\epsilon$ is the interaction strength, and $\sigma_{ij} = (\sigma_{i} + \sigma_{j})/2$. The interaction between  particle $i$ and the droplet interface, $U_{pw}(r_i,t)$,  is modeled as a soft repulsive harmonic potential
\begin{eqnarray}
U_{pw}(r_i,t) =
\left\{
\begin{array}{ll}
A_i \left(r_i - R(t) + \sigma_{i}/2\right)^2,
& r_i \ge (R(t)-\sigma_{i}/2), \\[6pt]
0, & r_i < (R(t)-\sigma_{i}/2),
\end{array}
\right.
\label{eq4}
\end{eqnarray}
where  $A_i$ is the spring constant. 

\begin{figure}[b]
\includegraphics[width=0.48\textwidth]{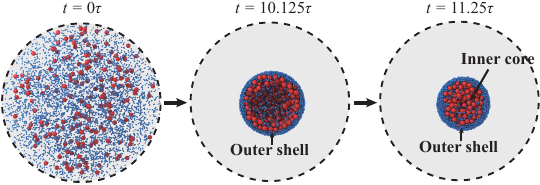}
\caption{\label{fig1_1}Typical simulation snapshots showing cross-sectional views  of the time evolution of particles confined within a droplet. As drying progresses, the initially well-mixed system gradually develops a pronounced stratified structure: smaller particles preferentially accumulate near the droplet surface, while larger particles become  enriched in the inner core. The configurations shown correspond to a size ratio $\sigma_l/\sigma_s=3$, an evaporation speed or P\'eclet number $Pe_s = 100$, and initial packing fraction $\Phi(0)=0.02$, where particle motion is slow compared to the evaporation rate.}
\end{figure}

\begin{figure*}[!tbp]
  \centering
    \includegraphics[width=\textwidth]{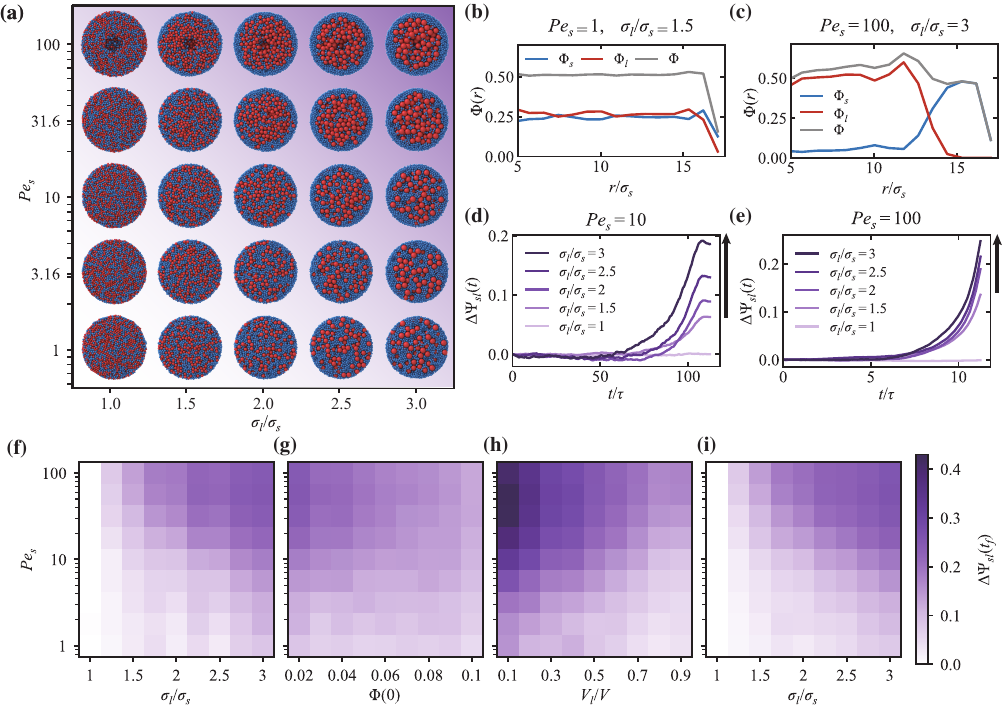}
  \caption{\label{fig1_2}{Brownian dynamics simulation results showing the effects of particle size ratio $\sigma_l/\sigma_s$, small-particle Péclet number $Pe_s$, initial packing fraction $\Phi(0)$, and large-particle volume fraction $V_l/V$ on stratification in binary mixtures during droplet drying. Here, $\sigma_s$ and $\sigma_l$ are the diameters of the small and large particles, respectively, and $V$ is the total particle volume. (a) Cross-sectional snapshots of the final particle configurations at initial packing fraction $\Phi(0)=0.02$ and equal volumes of small and large particles $V_s=V_l$. Each row corresponds to a fixed $Pe_s$, and each column to a fixed $\sigma_l/\sigma_s$; small and large particles are shown in blue and  red, respectively. (b,c) Final radial packing fraction profiles of the small particles $\Phi_s(r)$, the large particles $\Phi_l(r)$, and the total $\Phi(r)$, for (b) $Pe_s=1$, $\sigma_l/\sigma_s=1.5$ and (c) $Pe_s=100$, $\sigma_l/\sigma_s=3$. (d,e) Time evolution of the stratification parameter $\Delta\Psi_{sl}(t)$ for different size ratios at (d) $Pe_s=10$ and (e) $Pe_s=100$. (f--i) Heatmaps of the final stratification strength $\Delta\Psi_{sl}(t_f)$ in the $Pe_s$--$\sigma_l/\sigma_s$ parameter space for (f) $\Phi(0)=0.02$ and $V_s=V_l$; in the $Pe_s$--$\Phi(0)$ parameter space for (g) $\sigma_l/\sigma_s=3$ and $V_s=V_l$; in the $Pe_s$--$V_l/V$ parameter space for (h) $\sigma_l/\sigma_s=3$ and $\Phi(0)=0.02$; and in the $Pe_s$--$\sigma_l/\sigma_s$ parameter space for (i) with twice the total particle volume used in (f),  $\Phi(0)=0.02$ and $V_s=V_l$. Here, $t_f$ denotes the final drying time. All four heatmaps use the same color scale.}}
\end{figure*}

We perform all simulations using the LAMMPS software package (29 Aug 2024)\cite{LAMMPS} . The unit of time is defined as $\tau=\sigma_s^2/D_s$, where $\sigma_s$ and $D_s$ are the diameter and diffusion coefficient of the smallest particle species in the system, respectively. The packing fraction for the system is defined as $\Phi(t)= \sum_g \Phi_g(t)= \sum_g N_g\sigma_g^3 /(8 R^3(t))$, where $R(t)$ denotes the droplet radius at time $t$, $g$ indexes the particle species, $N_g$ is the number of particles of species $g$, and $\sigma_g$ is the corresponding diameter. The initial packing fraction $\Phi(0)$ is varied to control the initial droplet size. Unless otherwise stated, the total volume of particles $V= \sum_g V_g\approx10467\sigma_s^3$. All particles are initially placed randomly inside the droplet while rejecting configurations in which any pair lies within the interaction cutoff. Unless otherwise stated, simulations proceed with a time step $dt=2.5\times10^{-5}\tau$ until $\Phi(t_f)\approx0.64$, where $t_f$ denotes the final time of drying. The droplet shrinking rate $\alpha$ in Eq.(\ref{eq1}) is controlled by $Pe_s$ through $\alpha=2 Pe_sD_s$, where $Pe_s$ characterizes the evaporation rate relative to the Brownian diffusion of the smallest particle species. For particles of other sizes, the diffusion coefficient is taken to be inversely proportional to the particle diameter, $D_g=(\sigma_s/\sigma_g)D_s$, following the Stokes-Einstein relation. The interaction strength between all particle pairs is set to $\epsilon=40k_B T$. The spring constant for the smallest particles is $A_s=25k_B T/\sigma_s^2$, and for the remaining species it scales with the square of the particle diameter as $A_g=(\sigma_g/\sigma_s)^2A_s$.

\subsection{Multi-particle collision dynamics}

To investigate the role of hydrodynamic interactions, we compare multi-particle collision dynamics (MPCD) simulations with explicit solvent particles with  Brownian dynamics (BD) simulations with an implicit solvent. The simulations use  comparable conditions, allowing us to evaluate how solvent-mediated hydrodynamic interactions influence stratification. We perform the MPCD simulations using the stochastic rotation dynamics (SRD) implementation in LAMMPS (version 29 Aug 2024)\cite{LAMMPS}.

In the MPCD simulations, the mass of colloids of species $g$ is  assigned assuming a fixed mass density $m_g=(\sigma_g/\sigma_s)^3m_s$. The integration time step is $dt=4.5\times10^{-4}\sqrt{m_s\sigma_s^2/k_BT}$, and the SRD velocity-rotation step is performed every  $dt_{\mathrm{srd}}=4dt$. The SRD fluid has thermal energy $k_BT$ and a mean free path $\lambda=0.023\sigma_s$. Each SRD particle has mass  $m_{\mathrm{srd}}=k_BT(dt_{\mathrm{srd}}/\lambda)^2$. We set the  SRD number density  to $5/a_0^3$, where $a_0=0.23\sigma_s$ is the SRD grid cell size. 

To reduce multiple collisions between an SRD particle and nearby colloids, we apply a scaling factor of $0.93$ to the colloid–SRD collision radius. Slip boundary conditions are used for colloid--SRD collisions, and the SRD collision grid is randomly shifted before each velocity-rotation step. These parameter choices balance the computational cost of  large-scale droplet simulations with the resolution needed to capture  solvent-mediated hydrodynamic effects. The colloid--colloid and colloid--wall interactions are as described  in Section \ref{BD}.

With these parameters, the equations in Ref.~\citenum{petersen2010mesoscale} give an estimated shear viscosity of $\eta\approx3.618\sqrt{m_s k_BT}/\sigma_s^2$ and a Schmidt number of $Sc\approx3.565$. MPCD simulations often have lower Schmidt numbers than real liquids because achieving very large values can be computationally expensive. However, the solvent can exhibit  liquid-like behavior when momentum transport is faster than mass transport and the relevant time scales remain separated, as discussed in Ref.~\citenum{PhysRevE.74.031402}. In this work, we focus on how hydrodynamic interactions affect stratification within the adopted model, rather than on quantitatively reproducing  the properties of a specific real solvent.  A free-diffusion test based on the mean-squared displacement gives $D_s\approx0.0564\sigma_s\sqrt{k_BT/m_s}$, and the measured colloidal thermal energy is $k_BT_c\approx0.926k_BT$.

As in Section \ref{BD}, the colloids are confined within a shrinking spherical boundary of radius $R(t)$. The droplet is embedded in a cubic simulation box of side length $d\approx2.4R(0)$ with periodic boundary conditions. While the colloids remain confined to the spherical droplet,  SRD particles can  cross the periodic boundaries and move throughout the simulation box. Initially, the colloids are placed randomly inside the droplet, rejecting configurations in which any pair of particles lies within the interaction cutoff. The SRD particles are then distributed randomly throughout the simulation box, excluding the volume occupied by the colloids. Before droplet shrinkage begins, the colloids and SRD solvent are equilibrated. The droplet is subsequently compressed using the  protocol described in Section \ref{BD}. 

In the corresponding BD simulations, we use the same particle diameters, number of particles of each species, time step $dt$, initial volume fraction $\Phi(0)$, interaction potentials, and  droplet-shrinking protocol as  in the MPCD simulations. We set the values of $k_BT_c$ and $D_s$ according to the corresponding  estimates obtained from the MPCD simulations, while the diffusion coefficient of each particle species is set according to the Stokes--Einstein scaling relation, $D_g=(\sigma_s/\sigma_g)D_s$. The initial colloidal configuration is taken directly from the MPCD simulation after system relaxation.

\section{Results and discussion}
\subsection{Binary system}

We first investigate the stratification behavior of a binary system consisting of large particles of diameter $\sigma_l$ and small particles of diameter $\sigma_s$, yielding a size ratio $\sigma_l / \sigma_s$. The Péclet number $Pe_s$ controls the evaporation speed relative to the diffusion of the small particles, while the Péclet number for the large particles scales as $Pe_l = (\sigma_l / \sigma_s)Pe_s$. Fig.~\ref{fig1_1} illustrates the temporal evolution of the particle distribution for the case $Pe_s \gg 1$ and $Pe_l \gg 1$, where evaporation occurs much faster than particle diffusion. Under these conditions, the system exhibits pronounced stratification: large particles preferentially accumulate in the  inner core, while small particles are enriched near the droplet surface. This stratification behavior is consistent with previous studies\cite{liu2019segregation} and has also been observed in planar drying films \cite{fortini2016dynamic,howard2017stratification}. 

Fig.~\ref{fig1_2}(a) illustrates the combined effects of the Péclet number $Pe_s$ and the size ratio $\sigma_l/\sigma_s$ on small-large particle separation at the final drying time $t_f$. These parameters   characterize the relative evaporation rate and the particle size disparity, respectively. As $Pe_s$ increases from $1$ to $100$ and the size ratio $\sigma_l/\sigma_s$ from $1$ to $3$,  stratification becomes progressively more pronounced, with the strongest size segregation observed at high $Pe_s$ and large $\sigma_l/\sigma_s$. This trend is also reflected in the species-resolved radial packing fraction profiles $\Phi_g(r)$, defined as 
\begin{equation}
\Phi_g(r) =
\frac{
    \displaystyle\sum_{i\in g} V_{i,\mathrm{shell}}(r,r+\Delta r)
}{
    \displaystyle\frac{4\pi}{3}
    \left[(r+\Delta r)^3-r^3\right]
},
\end{equation}
where the denominator is the volume of the spherical shell spanning radii 
$r$ and $r+\Delta r$, and $V_{i,\mathrm{shell}}(r,r+\Delta r)$ is the portion of particle $i$'s volume that lies within that shell. A particle overlapping multiple radial bins contributes its volume separately to each bin. We sum these contributions over all particles of species $g$ and divide by the shell volume to obtain  $\Phi_g(r)$. For $Pe_s=1$ and a size ratio of $\sigma_l/\sigma_s=1.5$ in Fig.~\ref{fig1_2}(b), the packing fraction profiles of the small and large particles largely overlap, indicating negligible stratification. By contrast, increasing the size ratio to $\sigma_l/\sigma_s=3$ and the Péclet number to $Pe_s=100$ results in pronounced stratification (Fig.~\ref{fig1_2}(c)). Small particles preferentially accumulate  near the surface, while large particles become enriched in the interior. This represents the strongest stratification among the conditions shown in Fig.~\ref{fig1_2}(a). Video M1 shows the time evolution of the particle distributions corresponding to Fig.~\ref{fig1_2}(b) and (c).

The average radial position of species $g$ is 
${\bar r}_g(t) = (1/N_g) \sum_{i\in g} r_i(t)$,
where $N_g$ is the number of particles of species $g$ and $r_i(t)$ is the radial position of particle $i$. We quantify the degree of stratification  by 
\begin{equation}
\Psi_{gg'}(t) =
\frac{
    \displaystyle
    \frac{1}{N_g}\sum_{i\in g} r_i(t)
    -
    \frac{1}{N_{g'}}\sum_{i\in g'} r_i(t)
}{
    \displaystyle
    \max_i\left[r_i(t)+\frac{\sigma_i}{2}\right]
}.
\label{eq7}
\end{equation}
We normalize the separation measure by the actual outer radius of the particle assembly, $\max_i[r_i(t)+\sigma_i/2]$, defined by the outermost particle surface. With this convention, $\Psi_{gg'}>0$ indicates enrichment of particle species $g$ near the outer region, whereas $\Psi_{gg'}<0$ indicates preferential accumulation of particle species $g$ near the interior.

\begin{figure*}[t]
  \centering
  \includegraphics[width=\textwidth]{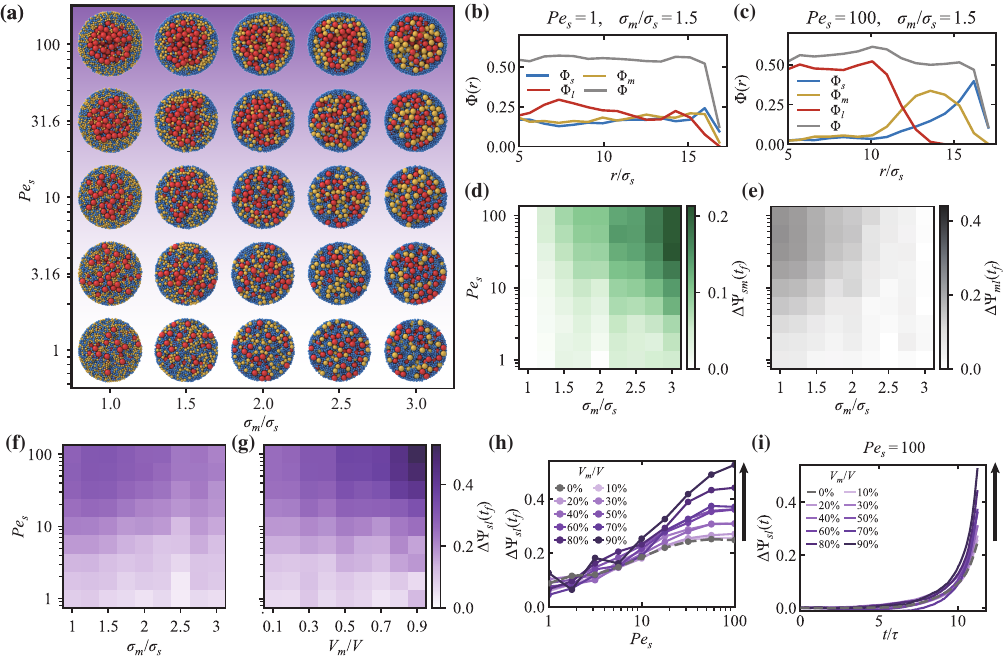}
  \caption{\label{Fig2_2} Simulation results illustrating the effects of the middle-particle diameter $\sigma_m$, Péclet number $Pe_s$, and middle-particle volume fraction $V_m/V$ on stratification in a ternary colloidal mixture during droplet drying. The small and large particle diameters are fixed at $\sigma_s$ and $\sigma_l=3\sigma_s$, respectively, and the initial packing fraction is $\Phi(0)=0.02$ throughout. (a) Cross-sectional snapshots of the final particle configurations for different $Pe_s$ and $\sigma_m/\sigma_s$. Each row corresponds to a fixed $Pe_s$, and each column to a fixed $\sigma_m/\sigma_s$. Small, middle, and large particles are shown in blue, yellow, and red, respectively. (b,c) Radial packing fraction profiles of the three species and their total at $\sigma_m/\sigma_s=1.5$ for (b) $Pe_s=1$ and (c) $Pe_s=100$. (d--f) Final pairwise stratification strengths in the $Pe_s$--$\sigma_m/\sigma_s$ parameter space: (d) small--middle, $\Delta\Psi_{sm}(t_f)$; (e) middle--large, $\Delta\Psi_{ml}(t_f)$; and (f) small--large, $\Delta\Psi_{sl}(t_f)$. In (a--f), the three species contribute equal particle volumes. (g) Final small--large stratification strength $\Delta\Psi_{sl}(t_f)$ in the $Pe_s$--$V_m/V$ parameter space at fixed $\sigma_m/\sigma_s=1.5$. The remaining particle volume is divided equally between the small and large species. (h) $\Delta\Psi_{sl}(t_f)$ as a function of $Pe_s$ for different $V_m/V$, including the binary mixture at $V_m/V=0$. (i) Time evolution of $\Delta\Psi_{sl}(t)$ at $Pe_s=100$ for the same compositions. Here, $V_m/V$ is the fraction of the total particle volume contributed by middle particles, and $t_f$ is the final drying time.}
\end{figure*}

Fig.~\ref{fig1_2}(d) shows the time evolution of $\Delta\Psi(t)=\Psi(t)-\Psi(0)$ for several values of the size ratio $\sigma_l/\sigma_s$ at fixed $Pe_s$. Subtracting $\Psi(0)$ removes the initial offset due to random initialization, so that $\Delta\Psi(t)$ captures the change in stratification during droplet shrinkage. For all size ratios shown, $\Delta\Psi(t)$ increases with time, indicating progressive enrichment of small particles near the outer shell. 
Comparing the curves in Fig.~\ref{fig1_2}(d), we find that larger size
ratios \ensuremath{\sigma_l/\sigma_s} correspond to higher
\ensuremath{\Delta\Psi(t)}, indicating stronger stratification. This
trend is consistent with Fig.~\ref{fig1_2}(a) and supports the use of
\ensuremath{\Delta\Psi(t)} as a measure of stratification strength. At
\ensuremath{Pe_s=100} (Fig.~\ref{fig1_2}(e)), larger size ratios again
lead to stronger stratification, although the dependence on size ratio is weaker than in Fig.~\ref{fig1_2}(d).
 
In Fig.~\ref{fig1_2}(f), we present a $9\times9$ heatmap of $\Delta\Psi(t_f)$ in the $Pe_s$--$\sigma_l/\sigma_s$ parameter space, where $t_f$ is the final drying time. The heatmap shows that stratification becomes more pronounced as both $Pe_s$ and $\sigma_l/\sigma_s$ increase. Similarly, Fig.~\ref{fig1_2}(g) shows  $\Delta\Psi(t_f)$ in the $Pe_{\mathrm{s}}$--$\Phi(0)$ parameter space. Stratification is stronger at higher $Pe_{\mathrm{s}}$ and lower $\Phi(0)$. At fixed total particle volume $V$, this corresponds to  faster evaporation and a larger initial droplet, both of which favor stratification. Fig.~\ref{fig1_2}(h) shows $\Delta\Psi(t_f)$ in the $Pe_s$--$V_l/V$ parameter space to examine how the volume fraction of large particles affects stratification. The heatmap shows stronger stratification at lower $V_l/V$ and higher $Pe_{\mathrm{s}}$, indicating that a large volume fraction of large particles is not necessary for strong stratification. To examine the effect of system size, we double the total particle volume in Fig.~\ref{fig1_2}(i) relative to Fig.~\ref{fig1_2}(f). The same trend persists: Stratification is stronger  at higher $Pe_s$ and larger $\sigma_l/\sigma_s$.

In conclusion, stratification is enhanced for faster evaporation (larger $Pe_s$), larger size ratios (larger $\sigma_l/\sigma_s$), a lower relative volume fraction of large particles (smaller $V_l/V$), and a lower initial volume fraction $\Phi(0)$ (or larger initial droplet size). Similar trends have  been reported in planar drying systems\cite{fortini2016dynamic, howard2017stratification, zhou2017cross}.
\begin{figure*}[!tbp]
  \centering
  \includegraphics[width=\textwidth]{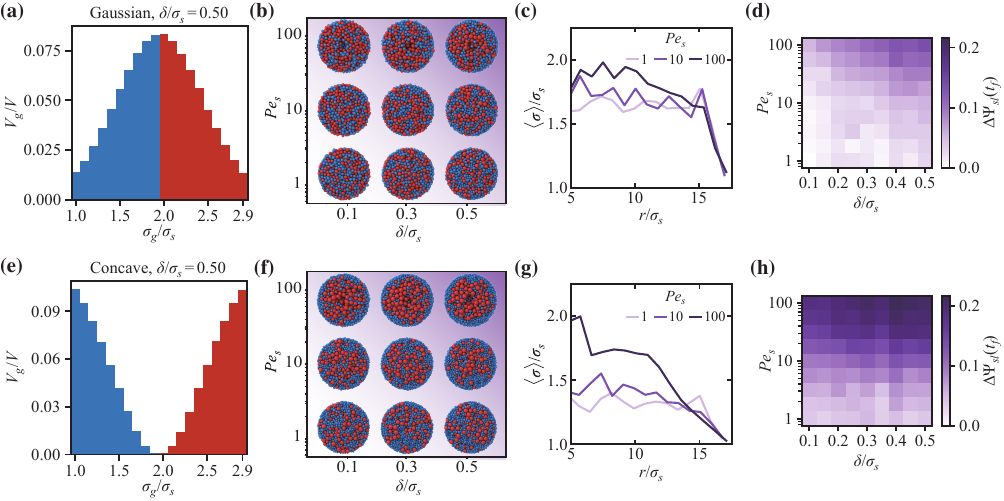}
  \caption{\label{fig3}Simulation results illustrating the effects of the particle-size distribution and Péclet number $Pe_s$ on stratification in drying polydisperse colloidal droplets. Particle diameters range from $\sigma_s$ to $2.9\sigma_s$, and both volume-weighted distributions are defined about $\sigma_{av}=1.95\sigma_s$. The initial packing fraction is fixed at $\Phi(0)=0.02$. (a,e) Volume fraction $V_g/V$ contributed by each particle species $g$ for the Gaussian and concave distributions, respectively, at $\delta/\sigma_s=0.5$. The parameter $\delta$ sets the width of the Gaussian distribution or the central depletion in the concave distribution. Blue and red denote particles smaller and larger than $\sigma_{av}=1.95\sigma_s$, respectively. (b,f) Cross-sectional snapshots of the final configurations for $\delta/\sigma_s=0.1$, $0.3$, and $0.5$ (columns) and $Pe_s=1$, $10$, and $100$ (rows). (c,g) Radial profiles of the mean particle diameter $\langle\sigma\rangle$ at $\delta/\sigma_s=0.5$ for the indicated $Pe_s$ values. (d,h) Final stratification strength $\Delta\Psi_{sl}(t_f)$ in the $Pe_s$--$\delta/\sigma_s$ parameter space. For this measure, particles below and above $\sigma_{av}=1.95\sigma_s$ are classified as small and large, respectively.}
\end{figure*}

\subsection{Ternary system}
Motivated by previous studies of planar drying, we next investigate stratification in a ternary mixture. The small and large particle diameters are fixed at $\sigma_s$ and $\sigma_l=3\sigma_s$, respectively, while the middle particle diameter $\sigma_m$ is varied. Fig.~\ref{Fig2_2} shows how the Péclet number $Pe_s$ and the size ratio $\sigma_m/\sigma_s$ affect the resulting particle distributions. The snapshots in Fig.~\ref{Fig2_2}(a) suggest that three-layer stratification is most apparent at high $Pe_s$ when $\sigma_m/\sigma_s$ is between approximately 1.5 and 2.0. The radial packing fraction profiles support this observation. At $Pe_s=1$ and $\sigma_m/\sigma_s=1.5$, Fig.~\ref{Fig2_2}(b) shows nearly uniform profiles for all three species. By contrast, at $Pe_s=100$, the profiles in Fig.~\ref{Fig2_2}(c) show a distinct three-layer structure: small particles are enriched near the droplet surface, large particles are concentrated toward the core, and middle particles form a layer between them. Video M2 shows the time evolution of the particle distributions for the conditions shown in Fig.~\ref{Fig2_2}(b) and (c).

We quantify  stratification using $\Delta\Psi_{sm}$, $\Delta\Psi_{ml}$, and $\Delta\Psi_{sl}$ for the small--middle, middle--large, and small--large pairs, respectively. The heatmaps in Figs.~\ref{Fig2_2}(d) and \ref{Fig2_2}(e) show that small--middle stratification is stronger at higher $Pe_s$ and larger $\sigma_m/\sigma_s$, whereas middle--large stratification is stronger at higher $Pe_s$ and smaller $\sigma_m/\sigma_s$. These opposing trends are consistent with the changes in size disparity: increasing $\sigma_m$ increases the size disparity between small and middle particles but decreases it  between the middle and large particles. Figure~\ref{Fig2_2}(f) shows that small--large stratification is also strongest at higher $Pe_s$ and relatively small $\sigma_m/\sigma_s$. However, compared with the middle--large case, the region of strongest stratification shifts slightly toward larger $\sigma_m/\sigma_s$.

\begin{figure*}[!tbp]
  \centering
  \includegraphics[width=\textwidth]{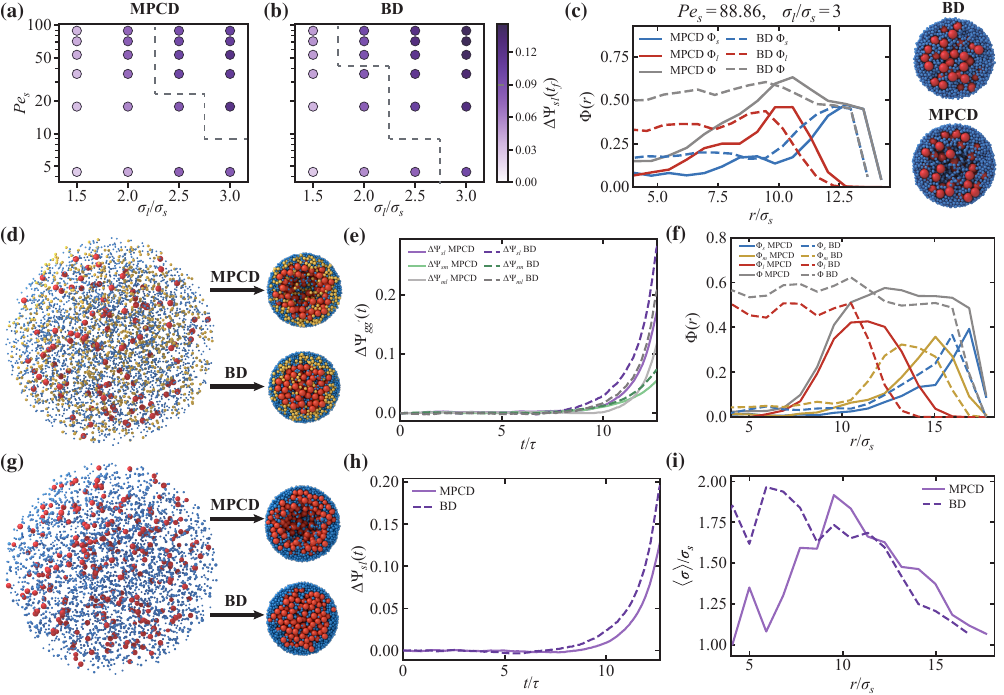}
  \caption{\label{fig4_1}Comparison of stratification in multi-particle collision dynamics (MPCD) and free-draining Brownian dynamics (BD) simulations for binary, ternary, and polydisperse systems: (a,b) Final stratification strength $\Delta\Psi_{sl}(t_f)$ in the $Pe_s$--$\sigma_l/\sigma_s$ parameter space for MPCD and BD, respectively, at fixed initial packing fraction $\Phi(0)=0.1$, final droplet packing fraction $\Phi(t_f)\approx0.69$, total particle volume $V\approx5233\sigma_s^3$, and relative volume fraction of large particles $V_l/V=0.4$. The gray dashed lines serve as visual guides enclosing state points with $\Delta\Psi_{sl}(t_f)>0.09$. Both panels use the same color scale. (c) Final radial packing fraction profiles for the corresponding binary systems at $Pe_s=88.86$ and $\sigma_l/\sigma_s=3$. (d) Cross-sectional snapshots of the common initial configuration and the final MPCD and BD configurations for the ternary system at $\Phi(0)=0.02$, $\Phi(t_f)\approx 0.64$, $Pe_s=88.86$, $\sigma_m/\sigma_s=1.5$, $\sigma_l/\sigma_s=3$ and $V_l\approx V_m\approx V_s$. (e) Time evolution of the pairwise stratification parameters $\Delta\Psi_{sl}(t)$, $\Delta\Psi_{sm}(t)$, and $\Delta\Psi_{ml}(t)$ for the ternary system. (f) Corresponding final radial packing fraction profiles. (g) Cross-sectional snapshots of the common initial configuration and the final MPCD and BD configurations for the polydisperse system with a volume-weighted concave size distribution at $\Phi(0)=0.02$, $\Phi(t_f)\approx 0.64$, $Pe_s=88.86$, and $\delta/\sigma_s=0.5$. (h) Time evolution of $\Delta\Psi_{sl}(t)$ for the polydisperse system, with particles classified as small or large according to whether their diameters are below or above $1.95\sigma_s$, respectively. (i) Corresponding final radial profiles of the mean particle diameter $\langle\sigma\rangle/\sigma_s$, calculated using 20 radial bins. In panels (c), (e), (f), (h), and (i), solid and dashed lines represent MPCD and BD results, respectively. Gray lines in panels (c) and (f) show the total packing fraction.}
\end{figure*}

We next examine how the volume fraction of the middle particles affects the small--large stratification. Fig.~\ref{Fig2_2}(g) shows a heatmap of $\Delta\Psi_{sl}(t_f)$ as a function of $Pe_s$ and $V_m/V$ at fixed $\sigma_m/\sigma_s=1.5$. Small--large stratification is stronger at higher $Pe_s$ and larger $V_m/V$. This trend is shown more directly in  Fig.~\ref{Fig2_2}(h), which presents $\Delta\Psi_{sl}(t_f)$ against $Pe_s$ for different values of $V_m/V$, including the binary mixture  $V_m/V=0$. Overall, the curves  shift upward as $V_m/V$ increases. Fig.~\ref{Fig2_2}(i) shows the time evolution of $\Delta\Psi_{sl}(t)$ at fixed $Pe_s=100$. Increasing $V_m/V$ also leads to stronger small--large stratification during drying. These results suggest that, under the conditions studied, adding middle-sized particles can enhance small--large stratification. In particular, this enhancement is observed when the  middle-particle diameter is closer to that of the small particles than to that of the large particles.

In conclusion, three-layer stratification in the ternary system is most pronounced at high $Pe_s$, particularly when the middle particles are   closer in size to the small particles than to  the large particles. The pairwise analysis shows that increasing $\sigma_m$ strengthens small--middle stratification while weakening middle--large stratification. At fixed $\sigma_m/\sigma_s=1.5$, increasing the volume fraction of the middle particles also enhances small--large stratification under the conditions studied here. Thus, both the size and volume fraction of the  middle species can be used to control the radial particle distribution.

\subsection{Polydisperse system}
For both binary and ternary systems, we observe clear stratification, with larger particles preferentially accumulating in the droplet interior and smaller particles near the droplet surface. This naturally raises the question of whether such behavior persists in more complex, polydisperse systems containing many particle species. 

To address this question, 

we first define a volume-weighted Gaussian distribution 
\begin{equation}
\frac{V_g}{V}
=
\exp\!\left(-\dfrac{(\sigma_g-\sigma_{av})^2}{2\delta^2}\right)
/
\displaystyle\sum_j
\exp\!\left(-\dfrac{(\sigma_j-\sigma_{av})^2}{2\delta^2}\right)
,
\end{equation}
where $\sigma_{av}=1.95\sigma_s$ is the center of the distribution, and $\delta$ is varied to control its width. The distribution is truncated to the interval $[\sigma_s,2.9\sigma_s]$, with particle diameters spaced at intervals of $0.1\sigma_s$. To characterize stratification, we categorize particles with diameters above  $\sigma_{av}$ as large  and those below it as small. Fig.~\ref{fig3}(a) shows an example of the particle-volume distribution across species for  $\sigma_{av}=1.95\sigma_s$ and $\delta=0.5\sigma_s$. Fig.~\ref{fig3}(b) shows typical particle configurations in the $Pe_s$--$\delta/\sigma_s$ parameter space. Stratification becomes more pronounced  at higher $Pe_s$ and larger $\delta/\sigma_s$. As in the binary system,   stratification strengthens with increasing $Pe_s$. It also becomes more pronounced as $\delta/\sigma_s$ increases, since a broader size distribution increases the disparity between particle sizes.  

For the polydisperse system, we also examine how the average particle diameter varies with radial distance from the droplet center to its boundary. We divide the droplet into concentric radial bins and calculate the average diameter in each bin 
\begin{equation}
\langle\sigma\rangle(r) =
\frac{1}{N(r,r+\Delta r)} \sum_{i \in [r,r+\Delta r]} \sigma_i,
\label{eq8}
\end{equation}
where the sum runs over all particles whose radial coordinate falls within $[r,r+\Delta r]$, and $N(r,r+\Delta r)$ is the number of particles in that interval. The results are shown in Fig.~\ref{fig3}(c), where the profiles correspond to different values of $Pe_s$. The average particle diameter generally decreases with increasing radial distance $r$. At higher $Pe_s$, particles in the interior have a larger mean particle diameter, and the average diameter decreases  more sharply toward the droplet surface, indicating stronger radial size segregation. This trend is further quantified in Fig.~\ref{fig3}(d), which shows a  heatmap  of $\Delta\Psi(t_f)$  in  the $Pe_s$--$\delta/\sigma_s$ parameter space. The heatmap confirms  that  stratification is strongest  at high $Pe_s$ and large $\delta/\sigma_s$.

To increase the particle size disparity, we introduce a volume-weighted concave distribution
\begin{equation}
\begin{aligned}
V_g/V
&= \left[1-\exp\!\left(
-\frac{(\sigma_g-\sigma_{av})^2}{2\delta^2}
\right)\right] \\
&\quad/
\left\{
\sum_j
\left[1-\exp\!\left(
-\frac{(\sigma_j-\sigma_{av})^2}{2\delta^2}
\right)\right]
\right\}.
\end{aligned}
\end{equation}
An example of this distribution is shown in Fig.~\ref{fig3}(e). It allocates a larger fraction of particle volume to the lower and upper ends of the size range, thereby increasing  the size disparity.  Fig.~\ref{fig3}(f) shows distinguishable stratification at high $Pe_s$ for all values of $\delta/\sigma_s$ considered. Consistent with this, the profiles of $\langle\sigma\rangle(r)$ in Fig.~\ref{fig3}(g)  decrease more steeply with increasing radial distance at high $Pe_s$. The heatmap in Fig.~\ref{fig3}(h) likewise shows stronger stratification  than that in Fig.~\ref{fig3}(d), as reflected by its darker shading.

In summary, polydisperse systems also exhibit radial stratification, with smaller particles preferentially accumulating near the droplet surface. Stratification becomes stronger with increasing $Pe_s$ and depends on the particle-size distribution. In particular, the concave size distribution produces stronger stratification, especially at larger $\delta/\sigma_s$, because it places more particle volume at the extremes of the size range and thus increases the size disparity. This is consistent with the binary-system results, which show that greater size disparity promotes stratification. Since polydisperse particles are more common in practice, these findings may help guide  control of particle distributions within supraparticles. Video M3 shows the time evolution of the particle distributions for the Gaussian and concave size distributions.

\subsection{Effects of hydrodynamic interactions on stratification}
In this section, we perform multi-particle collision dynamics (MPCD) simulations and compare the results with those obtained from Brownian dynamics (BD) simulations. We first examine the binary system, shown in Fig.~\ref{fig4_1}(a) for MPCD  and Fig.~\ref{fig4_1}(b) for BD simulations. To observe stratification within a manageable simulation time, we use an initial packing fraction of $\Phi(0)=0.1$, a total particle volume of $V\approx5233\sigma_s^3$, and a large-particle volume fraction of $V_l/V=0.4$. Across all state points considered,  BD yields higher values of  $\Delta\Psi_{sl}(t_f)$ than MPCD, indicating stronger stratification. The gray dashed lines mark the boundary of the region where $\Delta\Psi_{sl}(t_f)>0.09$ and serve as a visual guide. This region is substantially larger for the BD simulations, suggesting  that the free-draining BD model overestimates the stratification strength relative to MPCD under the conditions studied. 

The difference  between the two simulation methods is also reflected in the radial packing fraction profiles shown in Fig.~\ref{fig4_1}(c). At $Pe_s=88.86$ and $\sigma_l/\sigma_s=3$, both MPCD and BD exhibit clear  stratification, with small particles enriched near the droplet surface and large particles preferentially concentrated in the interior. However, the total packing fraction profiles, shown by the gray lines,  differ substantially.  The MPCD profile increases more steeply toward the surface, whereas the BD profile remains comparatively uniform. Similar differences have been reported in previous studies\cite{jee2026stratification,kundu2025exploring}. 

These differences may arise from solvent backflow and other hydrodynamic effects, which modify particle redistribution during rapid drying. In our simulations, the stronger accumulation of particles near the surface in MPCD is accompanied by  weaker small--large separation, as reflected by a lower $\Delta\Psi_{sl}(t_f)$.

For the ternary system, we perform simulations with $\Phi(0)=0.02$, $Pe_s=88.86$, $\sigma_m=1.5\sigma_s$, and $\sigma_l=3\sigma_s$. Fig.~\ref{fig4_1}(d) presents the MPCD and BD results, which start from the same initial configuration. Three-layer stratification is visible in both cases, although the MPCD configuration shows a particle-depleted region near the droplet center. Fig.~\ref{fig4_1}(e) shows the time evolution of $\Delta\Psi(t)$ for the small--large ($sl$), small--middle ($sm$), and middle--large ($ml$) pairs. For all three pairs, the BD simulations (dashed lines) exhibit stronger stratification than the MPCD simulations (solid lines) during the later stages of drying, suggesting that BD tends to overestimate stratification  under these conditions.  The corresponding radial packing fraction profiles in Fig.~\ref{fig4_1}(f)  also show ternary stratification in both simulations. As in the binary system, the total packing fraction increases more steeply toward the surface in MPCD simulations. 

The polydisperse system shows similar behavior as presented in Fig.~\ref{fig4_1}(g): MPCD  produces a hollow structure relative to BD, although both simulations start from the same initial configuration. In Fig.~\ref{fig4_1}(h), we again classify particles as small or large using $\sigma=1.95\sigma_s$ as the size threshold. The time evolution of $\Delta\Psi_{sl}(t)$ again indicates  stronger stratification in BD. Figure~\ref{fig4_1}(i) compares the radial profiles of $\langle\sigma\rangle/\sigma_s$. In  MPCD, the occupied bins near the droplet center have a lower average particle diameter.

In summary, under the conditions considered, free-draining BD simulations tend to overestimate stratification relative to  MPCD. MPCD  also tends to produce a particle-depleted interior, a feature previously attributed to solvent backflow \cite{kundu2025exploring}. Nevertheless, stratification and size segregation persist in all three systems, when hydrodynamic interactions are included, albeit at reduced strength  compared with BD. Video M4 shows representative  BD--MPCD comparisons for all three systems.

\section{Conclusions}
We performed Brownian dynamics simulations on binary, ternary, and polydisperse mixtures of colloidal particles within a drying spherical droplet. 
For the binary system, we demonstrate that stratification can occur, with larger particles preferentially accumulating in the interior and smaller particles near the surface, consistent with previous observations reported in Ref.~\citenum{liu2019segregation}. 
By systematically varying the Péclet number of the smallest particles, $Pe_s$, which controls the relative evaporation rate, together with the particle size ratio $\sigma_l/\sigma_s$,  initial particle fraction $\Phi(0)$ and volume fraction of large particles $V_l/V$, we find that stratification increases with increasing $Pe_s$ and $\sigma_l/\sigma_s$, but weakens with increasing $\Phi(0)$ and $V_l/V$.

In ternary systems, clear three--layer stratification emerges at high $Pe_s$  when the middle particle size $\sigma_m$ is appropriately chosen. For $\sigma_l=3\sigma_s$, this structure is most apparent when $\sigma_m/\sigma_s$ is approximately 1.5--2.0:  large particles accumulate in the central region, small particles preferentially reside near the surface, and particles of intermediate size occupy the region in between. This hierarchical stratification is again consistent with observations reported for planar geometries~\cite{fortini2017stratification}. Our results further show that, when the middle particles are closer in size to the small particles than to the large ones, they can  enhance small--large segregation, and that this segregation becomes stronger as the volume fraction of middle particles $V_m/V$ increases. Thus,  adding middle-sized particles of an appropriate diameter may provide a means of enhancing stratification in binary mixtures.

For polydisperse systems, we find that  the mean particle size   generally decreases with increasing radial distance under  high-$Pe_s$ conditions. 
Moreover, the magnitude of this radial size gradient depends on the particle size distribution. Specifically, distributions with larger size disparities exhibit a more pronounced decrease in mean particle size with increasing $r$. Since particle sizes are often polydisperse in real systems, these results suggest that size segregation can be controlled by adjusting the evaporation rate and particle size distribution.

Including hydrodynamic interactions generally leads to weaker stratification  and more pronounced particle depletion in the droplet interior than in free-draining Brownian dynamics simulations. These effects are observed in the binary, ternary, and polydisperse systems studied here. Nevertheless, stratification persists in all three systems under the conditions tested. The MPCD solvent used here has a Schmidt number lower than that typical of real liquids, reflecting a compromise between computational cost and solvent dynamics. Accordingly, we interpret the MPCD--BD  comparisons as qualitative assessments of  hydrodynamic effects within the adopted model. Although the BD simulations use the same initial configurations as the MPCD simulations and are matched to them in temperature and small-particle diffusivity, the two methods do not provide identical descriptions of particle dynamics.

The stratification phenomena observed here have important implications for the fabrication of hierarchically structured materials through  evaporation-driven self-assembly. In particular, droplet-confined drying offers a simple and versatile route to engineer radial composition and size gradients in colloidal assemblies. Our results therefore provide deeper insight into supraparticle formation and the regulation of the internal structural organization. Such understanding is essential for the rational design of functional supraparticles, with potential applications in catalysis~\cite{li2017hierarchically,liu2019tuning,hou2019formation}, gas adsorption and sensing~\cite{fujiwara2023fast,fang2009single}, drug encapsulation and delivery~\cite{rial2014bioinspired,zhao2015microfluidic}, photonics~\cite{manoharan2001photonic,park2014full,vogel2015color}, and energy production and storage~\cite{koo2008nano,li2014mesoporous}.

\begin{acknowledgments}
This work is supported by the Industrial Partnership Programme Fundamental Fluid Dynamics Challenges in Inkjet Printing (Phase II): Physicochemical Hydrodynamics of Droplets (KICH2.V4C.20.001) of The Netherlands Organisation for Scientific Research (NWO), cofinanced by Canon Production Printing Netherlands B.V.
\end{acknowledgments}

\section*{Data Availability Statement}
The data and calculations that support the study are available on reasonable requests. 
\appendix

\nocite{*}
\bibliography{aipsamp}

\end{document}